\documentclass[osajnl,twocolumn,showpacs,superscriptaddress,10pt]{revtex4-1} 
\usepackage{amsmath,amssymb,graphicx}
\usepackage{color}

\def\figref#1{Fig.~(\ref{#1})}
\def\eqnref#1{Eq.~(\ref{#1})}
\begin{document}

\title{Generation of microwave radiation by nonlinear interaction of a high-power, high-repetition rate, 1064-nm laser in KTP crystals}
\author{A. F. Borghesani}
\email[]{armandofrancesco.borghesani@unipd.it}
\affiliation{CNISM unit, Department of Physics \& Astronomy, University of Padua, 
 Italy}\affiliation{Istituto Nazionale di Fisica Nucleare, Sezione di Padova, 
 Italy} 
\author{C. Braggio}\affiliation{Istituto Nazionale di Fisica Nucleare, Sezione di Padova, 
 Italy} 
\affiliation{Department of Physics \& Astronomy, University of Padua, 
 Italy}\
\author{G. Carugno}
\affiliation{Istituto Nazionale di Fisica Nucleare, Sezione di Padova, 
 Italy} 

\begin{abstract}
We report measurements of microwave (RF) generation in the centimeter band accomplished by irradiating a nonlinear KTiOPO$_4$ (KTP) crystal with a home-made, infrared laser at $1064\,$nm as a result of optical rectification (OR). 
The laser delivers pulse trains of duration up to $1\,\mu$s. Each train consists of several high-intensity pulses at an adjustable repetition rate of approximately $ 4.6\,$GHz. The duration of the generated RF pulses is determined by that of the pulse trains.
We have investigated both microwave- and second harmonic (SHG) generation as a function of the laser intensity and of the orientation of the laser polarization with respect to the crystallographic axes of KTP.
\end{abstract}
\ocis{(350.4010) Microwaves, (160.4330) Nonlinear optical materials, (160.2100) Electro-optical materials, (140.4050) Mode-locked lasers}
\maketitle

Optical heterodyning is at the basis of several techniques to produce microwave signals~\cite{bridges1972,yao2009,Khan2010}.
Two adjacent laser lines are brought to beat in a nonlinear crystal to generate signals at the difference frequency, which lies in the microwave domain. Authors have demonstrated generation of beat notes at frequencies from a few GHz up to the THz band~\cite{niebuhr1963,Lengfellner1987}. 

Optoelectronic oscillators have also been developed that generate spectrally pure microwave signals by modulating continuous laser light in interferometric devices~\cite{leader2002}, whose performance is limited by the bandwidth of the state-of-the-art photodetectors~\cite{neyer1982,yao1996}.

In this paper we report about a novel technique of microwave and mm-wave signal photonic generation. The method is demonstrated at $\approx 4.6\,$GHz but can be extended up to a few hundreds GHz ~\cite{keller2002,keller2003}. It is based on the fast response of a second order nonlinear crystal to a pulse train delivered by a high-intensity, mode-locked laser system, whose repetition rate $f$ is in the RF domain. The irradiation of such a crystal with a train of short laser pulses produces a time-dependent polarization in the crystal as a consequence of optical rectification (OR)~\cite{bass1962}. This process gives origin to the emission of microwave radiation that can be transferred to any receivers, either a cavity or a waveguide,  without the bandwidth limitation of photodetection. 

OR has also been used to produce subpicoseconds THz pulses by using ultrashort laser pulses in a number of nonlinear crystals~\cite{zhang1992,graf2000}.

We used a KTiOPO$_4$ (KTP) crystal because its non centrosymmetric, orthorombic crystal structure belonging to the $2mm$ point symmetry group~\cite{yariv2007} endows it with a strong second order nonlinearity. Its electro-optic coefficients are well known~\cite{bierlein1986,boulanger1994,pack2004} and it is, thus, very well suited to characterize our technique to produce RF radiation.

\begin{figure}[t!] 
  \centering
  \includegraphics[width=\columnwidth]{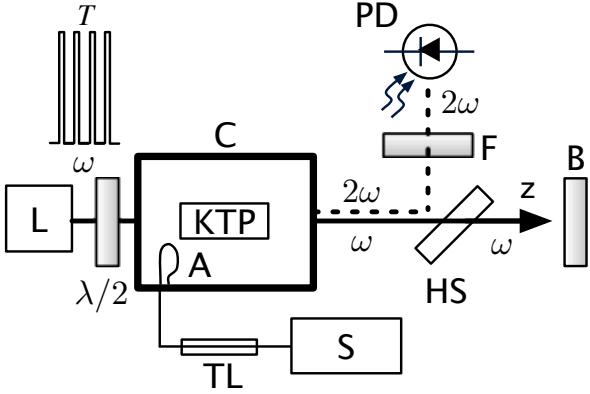}
  \caption{\small 
  Experimental setup. L = laser, $T$~=~pulse train, $\lambda/2$ = retarding plate, C = cavity, KTP~=~crystal, A = Antenna, TL = transmission line, S = scope, HS = harmonic separator, F = filter, PD = photodiode, B~=~bolometer. \label{fig:bd}}
  \end{figure}

 In \figref{fig:bd} a scheme of the experimental setup is shown.
The KTP crystal used (EKSMA Optics) is a parallelepiped  $4\times 4\times 10 \,$mm$^3$ in size whose long axis is aligned in the direction $z$ of the laser propagation. The crystal is cut in a way that its optical axis $z^\prime$ forms an angle $\beta\approx 59^\circ$ with $z,$ necessary for phase matching purposes in a previous experiment. 
It is mounted in the center of a rectangular RF cavity (C) designed so as to sustain a TE$_{111}$ mode.
Its resonance frequency can be tuned to the laser pulses repetition rate that can be varied in the range $4.4\,$GHz to $4.7\,$GHz.

The crystal can be rotated about the $z$ axis to maximize the coupling of the RF emission with the cavity mode. The quality factor of the loaded cavity is $Q\approx 10^3 .$ A critically coupled antenna (A) is used to pick up the RF signal that is directly fed to the $50\,\Omega $ input of the oscilloscope (S).
 
 The infrared ($\lambda=1064\,$nm) laser (L), described elsewhere~\cite{braggio2011}, is set to deliver one $400\,$ns or more-long pulse train every second. The pulse width is $\tau\approx 12 \,$ps and each train contains $\approx 2000$ of them, the pulse repetition rate being set at $f= 4.6\,$GHz.
 \noindent
 In this way, a restricted-band frequency comb~\cite{cundiff2003} is generated with mode frequencies $f_L\pm m f,$ where $f_L = c/\lambda\approx 2.82\times 10^{14}\,$rad/s is the laser frequency.
Because of the gaussian shape and duration of the individual pulses, the highest overtone of nonnegligible amplitude has index $m_\mathrm{max}\approx 20,$ so that  $m_\mathrm{max}f\ll f_L,$ and all the frequencies in the laser spectrum are in the optical region~\cite{boyd1971}.

 The maximum laser intensity value is limited to $I\simeq 130 \,$MW/cm$^2 $ in order to keep $I$ well below the damage threshold ($> 500\,$MW/cm$^2$ for $10\,$ns long pulses). $I$ can be reduced by inserting calibrated neutral density filters in the laser path. We note that the lower limit to the laser intensity is set by the requirement that it forces the nonlinear response of the crystal. We experimentally observed microwave generation by KTP with $I$ as low as $0.2\,\mbox{MW/cm}^{2}.$
 
The laser polarization can be rotated relative to the crystal axes by means of a $\lambda/2$ wave plate mounted on a rotating goniometer. 
The laser beam has an ellissoidal gaussian profile with semiaxes $a\approx1.5\, $mm and $b\approx 1.8\,$mm, respectively, and is fully projected onto the  antireflection coated, entrance face $(x,y)$ of the crystal.
The light exiting the crystal output face contains the contribution due to SHG and to the pump laser. The SH is picked out by a combination of a harmonics separator (HS) and a bandpass filter (F) and is measured by a photodiode (PD) whose output voltage $V_G$ is proportional to its intensity. A bolometer (B) is used to monitor the laser stability.
   
\begin{figure}[t!]
\begin{center}
 \includegraphics{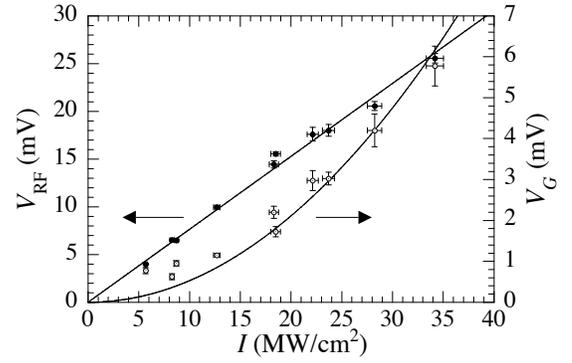}\caption{\small Microwave amplitude 
 $V_\mathrm{RF} $ (left scale) and SH intensity 
 $V_G$ (right scale)  vs laser intensity 
 $I$ for at fixed $\theta =115^\circ .$ 
\label{fig:VrfANDVGvsIlaser}}
 \end{center} 
 \end{figure}

In~\figref{fig:VrfANDVGvsIlaser} we show the amplitude $V_\mathrm{RF}$ of the microwave signal and the SH intensity $V_G$ as a function of the laser intensity $I$ for a fixed position of the $\lambda/2$ plate. In our experiment, the efficiency of SHG is of a few \%. The presence of SH is a clear sign of a quadratic nonlinear effect. 
Actually, 
 $V_G \propto I^2,$ the proportionality constant depending on the second-order susceptibility tensor for the mixing of two identical frequencies $d_{ijk}^{2\omega}.$
    
By contrast, the antenna signal is proportional to the time derivative of the microwave polarization field that depends on the quasi static polarization induced by OR. 
Thus, the amplitude of the RF signal is directly proportional to the laser intensity, $V_\mathrm{RF}\propto I,$ the proportionality constant depending on $d_{ijk}^0.$
 The experimental data confirm these expectations.
 
\begin{figure}[b!]
\begin{center}
 \includegraphics{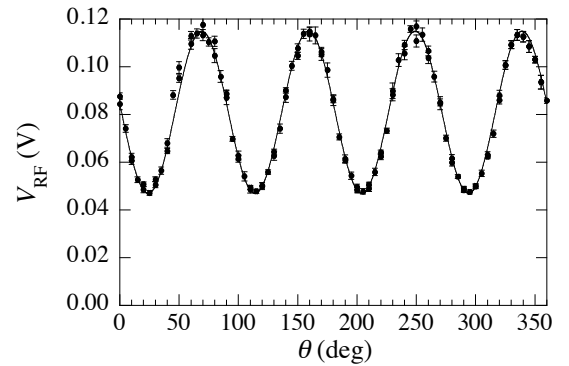}
 \caption{\small Microwave signal amplitude 
 $V_{\mathrm{RF}}$ vs 
 rotation angle $\theta$ of the $\lambda /2$ plate for $I\approx 100\, $MW/cm$^2 . $ 
  \label{fig:VrfFFtvsTheta}
}
\end{center} 
 \end{figure}

 To further verify that the microwave radiation is produced by OR, we measured the dependence of $V_\mathrm{RF}$ on the angle $\theta $ by which the direction of the laser polarization is rotated by the $\lambda/2$ plate. 
 The results are displayed in~\figref{fig:VrfFFtvsTheta}. 
 $V_\mathrm{RF}$ shows a marked 4-fold periodicity and can be accurately fitted to the following form
  \begin{equation}  
  V_\mathrm{RF}= V_0 + V_4 \cos{\left[4\left(\theta -\gamma_4\right)\right]}
  \label{eq:vrftheta} 
  \end{equation} 
with $V_0=80.9\,$mV, $V_4=33.9\,$mV. $\gamma_4=69^\circ$ has no physical relevance.

As the antenna signal is proportional to the first time derivative of the microwave field, the angular dependence of the nonlinear polarization can be calculated from $d_{ijk}.$ 
The low-frequency crystal polarization $P(t)$ closely follows the envelope $\vert E_{0j}^\omega (t)\vert^2$ of the electric field $E_j (t)=E_{0j}^\omega(t)\cos{\omega t}$ of the optical pulse train~\cite{graf2000}, where $\omega=2\pi f_L$ is the angular frequency of the 
laser. The Fourier spectrum of $P(t)$ contains the fundamental harmonic at the microwave frequency $f$ plus several, higher-order harmonics. 
 The resonant cavity acts as narrow band-pass filter that picks out the amplitude $P_0$ of the fundamental that
 can be written in the frame of reference of the crystallographic axes $(x^\prime,\, y^\prime,\, z^\prime)$ as
\begin{equation}
P_{0i}=A
d^{\prime 0}_{ijk}E_{0j}^\omega E_{0k}^\omega\qquad{(i,\,j,\,k=x^\prime,\, y^\prime,\, z^\prime)}
\label{eq:p0i}  
\end{equation} 
The  convention of summation over repeated indices is observed.
The constant $A$ depends on the actual shape of the individual pulses in the train.
$d^{\prime 0}_{ijk}$ are the elements of the nonlinear susceptibility tensor and the prime means that they are referred to the crystallographic system. For KTP crystals, only 5 of its elements are non null~\cite{yariv2007}. The crystallographic system is rotated with respect to the
 laboratory frame of reference so that the crystal axis $z^\prime$ forms an angle $\beta$ with the laser propagation direction $z.$ Thus, the polarization components in~\eqnref{eq:p0i} must be expressed in the laboratory frame. 

The laser light is linearly polarized with components $E_{0x}^\omega=E_{0y}^\omega=E_0$ in the laboratory. A rotation of the $\lambda/2$ plate by an angle $\theta$ rotates the laser polarization by $2\theta.$ Thus, the components of the field incident on the entrance face of the crystal are $E_x^\omega=E_0\cos 2\theta$ and $E_y^\omega=E_0\sin2\theta.$ 

The elements of the nonlinear second-order susceptibility tensor for OR, $d_{ijk}^{\prime 0},$ and for SHG, $d_{ijk}^{\prime 2\omega},$ could in principle be different.
However, we argue that they are nearly the same. In fact, as in our case there is only one, strong, laser source, the nonlinear, quadratic, electro-optic tensor is of electronic origin~\cite{boyd1971}. In the case of resonant mixing, if the driving frequencies are in the optical domain whereas the difference frequency is below any lattice resonances, there is no contribution of the lattice to  $d_{ijk}^\prime$~\cite{garrett1968}.
Moreover, the tensor elements can be written as $d_{ijk}^{\prime \omega_2\pm\omega_1}\propto \chi^{(1)}_e\left(\omega_2\pm\omega_1\right) \chi^{(1)}_e(\omega_1) \chi{(1)}_e(\omega_2),$ 
where the $\chi^{(1)}_e$'s are the linear electronic susceptibilities. They are related to the dielectric constant $\epsilon$ of the material and, thus, to the refraction index~\cite{miller1964}. 
It is reported that the static dielectric constants $\epsilon_{jj}(0)$ do not differ by more than a few \% from the square of the refraction indices at optic frequencies 
$n^2_{jj}(\omega)$~\cite{reshak2010}. For these reasons, we assume $d^{\prime 0}_{ijk} \simeq d^{\prime 2\omega}_{ijk}$ and drop either superscripts. 

The effective, non vanishing elements of the nonlinear second-order susceptibility tensor are thus expressed in the laboratory frame as
\begin{eqnarray}
  d_{15}&=&2 b d^\prime_{15} \cos 2\theta\sin2\theta\nonumber \\
  d_{24}&= &2 ab d^\prime_{24} \sin^2 2\theta\nonumber \\
  d_{31}&= &d^\prime_{31} \cos^2 2\theta\nonumber \\  
   d_{32}&= &d^\prime_{32} a^2\sin^2 2\theta\nonumber \\  
 d_{33}&=&d^\prime_{33} b^2 \sin^2 2\theta\label{eq:dprime}
 \end{eqnarray}   
in which $a=\cos\beta$ and $b=\sin\beta.$

 $V_\mathrm{RF}$ 
 is now given by the projection of the 
induced low-frequency polarization onto the direction of the electromagnetic mode of the cavity. It is easy to show that
\begin{equation}
V_\mathrm{RF} = B E_0
^2g(\theta)=C I g(\theta)\label{eq:Vrfgt}
\end{equation}
where $I=c\epsilon_0E_0^2$ is the laser intensity.  $\epsilon_0$ is the vacuum permittivity.
$B=B(f,\omega)$ and $C=B/c\epsilon_0$ are constant at fixed RF frequency and laser pulsation $\omega.$ $B$ accounts for many parameters such as effective interaction volume, antenna efficiency, and so on. 
The function 
$g(\theta)$ is given by
\begin{equation}
  g(\theta)= g_x d_{15}+ g_y  d_{24} +g_z\left(d_{31} +d_{32} + d_{33} 
  \right)\label{eq:gtheta}
\end{equation}
The direction cosines $g_x,\, g_y,$ and $g_z$ of the induced, 
nonlinear polarization field relative to the cavity mode polarization are unknown and are obtained by a fit to the experimental 
data.

\eqnref{eq:Vrfgt} explains both the results for $V_\mathrm{RF}$ in \figref{fig:VrfANDVGvsIlaser} and in \figref{fig:VrfFFtvsTheta}. 
The $V_\mathrm{RF}$ data in \figref{fig:VrfANDVGvsIlaser} are measured at fixed angle $\theta=\tilde \theta.$ So, $g(\tilde \theta)$ is a constant and $V_\mathrm{RF}$ turns out to be directly proportional to the laser intensity $I.$

The data shown in \figref{fig:VrfFFtvsTheta} are obtained at constant $I,$ so they display the behavior of $g(\theta).$ It can be shown that, by expanding the trigonometric functions,~\eqnref{eq:gtheta} can be cast in the form given by \eqnref{eq:vrftheta}.
By inserting the known values of the nonlinear electro-optic coefficient of KTP~\cite{pack2004}, the values of the direction cosines can be determined by comparison with the fit parameters in  
 \eqnref{eq:vrftheta}. If $g_x=-3.72\times 10^{-2},$ $g_y=-0.911,$ and $g_z=0.411$ are chosen, \eqnref{eq:gtheta} accurately fits the experimental data.  

A further verification of the validity of \eqnref{eq:vrftheta} can be obtained by measuring $V_\mathrm{RF}$ as a function of the laser intensity $I$ for several angular positions of the $\lambda/2 $ plate. 
 \begin{figure}[b!]
 \centering
 \includegraphics{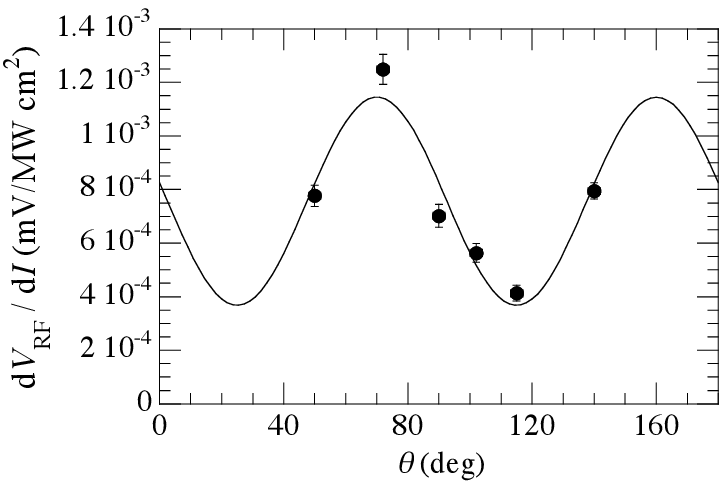}
\caption{\small 
Dependence of 
$\mathrm{d}V_{\mathrm{RF}}/\mathrm{d}I$ on the rotation angle $\theta $ of the $\lambda/2 $ plate.
\label{fig:slope}} 
 \end{figure}
From plots similar to \figref{fig:VrfANDVGvsIlaser}, we determined the slope of $V_\mathrm{RF} $ vs $I$ for some values of $\theta .$ The slope values are reported in \figref{fig:slope}. According to \eqnref{eq:Vrfgt}, the slope is proportional to $g(\theta)$ and the experimental results confirm this expectation.

\begin{figure}[t!]
\begin{center}
 \includegraphics{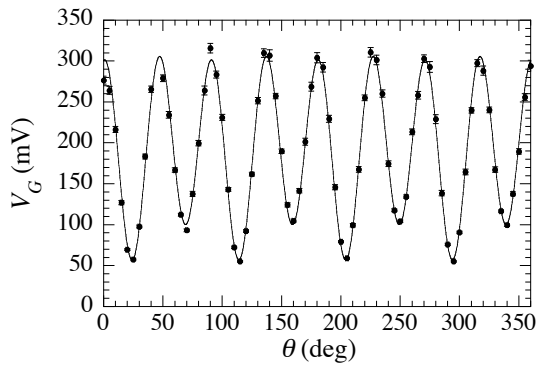}%
 \caption{\small Dependence of the SH intensity 
 $V_{G}$ on the rotation angle 
 vs $\theta$ of the $\lambda/2$ plate for $I\approx 74\,$MW/cm$^2.$ 
 \label{fig:VG}}
 \end{center}
 \end{figure}

\noindent SHG is also a consequence of the quadratic nonlinearity of the medium polarizability~\cite{yariv2007}. 
In \figref{fig:VrfANDVGvsIlaser} we have shown that the SH intensity quadratically depends on $I.$ 
In \figref{fig:VG} 
the dependence of the SH intensity on the direction $\theta$ of the pump laser polarization for $I\approx 74\,$MW/cm$^2 $ is shown.
The data show an 8-fold periodicity modulated by a 4-fold one and 
are fitted to a function of the form
\begin{equation}
  V_G = V_0 + V_8\cos{\left[8\left(\theta-\gamma_8\right)\right]} +V_4\cos{\left[4\left(\theta-\gamma_4\right)\right]}
\label{eq:vgv8v4}
\end{equation}
with $V_0= 190.8\,$mV, $V_8=112.2\,$mV, and $V_4=21.7\,$mV.
$\gamma_8=46.6^\circ$ and $\gamma_4=67.3^\circ$ have no physical relevance.
 
\noindent  The dependence of $V_{G}$ 
on the direction of the laser polarization is computed in a way similar to $V_{\mathrm{RF}}.$ By recalling that $V_{G}$ is proportional to the square of the second time derivative of the induced nonlinear polarization
and by recalling that the SH is nearly copropagating with the
laser along 
$z,$
$ V_G$ can be written as
\begin{equation}
  V_G=D\omega^4 I^2 q(\theta)  \label{eq:SHint}
\end{equation}
with
$
  q(\theta)= q_x 
  d_{15}^2 +q_y   d_{24}^2 
+q_z\left(d_{31}^2+d_{32}^2 
  + d_{33}^2
  \right).$ 
The constant $D $ depends on the interaction volume, on the photodetection efficiency, and so on. 
$q_x,$ $q_y,$ and $q_z$ 
 give the orientation of the crystal relative to the propagation axis accounting for a possible non perfect alignment of the geometrical $z$ axis of the crystal with respect to the laser propagation direction. The choice $q_x= 0.162,$ $q_y=-0.984,$ and $q_z=0.081$ gives an excellent fit to the data, as displayed as a solid line in \figref{fig:VG}.

In this Letter we have shown that microwave radiation in the centimeter band is optically produced by exploiting the nonlinear polarization properties of KTP crystals. The irradiation of a second-order nonlinear crystal with high-intensity laser pulses at a repetition rate in the microwave domain produces a modulation at the same frequency of OR, which is the source of RF radiation.  

The phenomenon we have described could be exploited as a new technique to  build flexible RF sources for applications, in which the RF pulse duration and frequency have to be tailored according to specific needs. To this goal, a characterization of such source including the determination of  the spectral density of the phase noise may be necessary. In this experiment, the limited duration of the 
RF pulses prevents its measurement.
However, we reasonably expect that in a CW system this feature is mainly determined by the frequency stability of the optical oscillator~\cite{millo2009}.
Moreover, 
 the bandwidth of the produced microwave signal is determined by both the pulse train  duration and the cavity $Q$ value.
 
This methodology can also be a useful tool to characterize the nonlinear electro-optic coefficients of crystals.

The authors thank E. Berto for technical assistance and acknowledge financial support by Istituto Nazionale di Fisica Nucleare within the MIR experiment.

    \clearpage\newpage

\end{document}